\documentclass[showkeys,superscriptaddress,nofootinbib,aps,floatfix,pra, onecolumn, 11pt]{revtex4-2} 
\usepackage{multirow}%
\usepackage{amsmath,amssymb,amsfonts}%
\usepackage{amsthm}%
\usepackage{mathrsfs}%
\usepackage[title]{appendix}%
\usepackage{subfigure}
\usepackage{graphics,graphicx}
\usepackage{url} 
\usepackage{hyperref}
\usepackage[a4paper, total={7in, 10in}]{geometry}
\usepackage[dvipsnames]{xcolor}
\usepackage{textcomp}%
\usepackage{booktabs}%
\usepackage{algorithm}%
\usepackage{algorithmicx}%
\usepackage{algpseudocode}%
\usepackage{listings}%
\usepackage{times}
\usepackage{nameref} 
\usepackage{tabularx}

\usepackage{epsfig}
\usepackage{verbatim}
\usepackage{calrsfs}
\usepackage{psfrag}
\usepackage{bbold}
\usepackage[english]{babel}
\usepackage{graphicx}
\usepackage{braket}
\usepackage{natbib}
\usepackage{lipsum}
\usepackage[section]{placeins} 

\usepackage{times}  
\usepackage{setspace}   
\usepackage[title]{appendix}
\usepackage{lineno}
\usepackage{xcolor}
\usepackage{xspace}
\usepackage{booktabs}
\usepackage{siunitx}
\usepackage{adjustbox}
\usepackage[normalem]{ulem}

\begin{document}
	\title{Socioeconomic disparities in mobility behavior during the COVID-19 pandemic in developing countries}
	
	\author{Lorenzo Lucchini$^{*, \dagger,}$ }
	\affiliation{Centre for Social Dynamics and Public Policy, Bocconi University, Milan, Italy}
	\affiliation{Institute for Data Science and Analytics, Bocconi University, Milan, Italy}
	\affiliation{World Bank Group, Washington, DC, USA}
	\affiliation{Fondazione Bruno Kessler, Trento, Italy}
	\author{Ollin D. Langle-Chimal$^*$}
	\affiliation{University of California at Berkeley, Berkeley, CA, USA}
	\affiliation{University of Vermont, Burlington, VT, USA}
	\affiliation{World Bank Group, Washington, DC, USA}
	\author{Lorenzo Candeago}
	\affiliation{World Bank Group, Washington, DC, USA}
	\author{Lucio Melito}
	\affiliation{World Bank Group, Washington, DC, USA}
	\author{Alex Chunet}
	\affiliation{World Bank Group, Washington, DC, USA}
	\author{Aleister Montfort}
	\affiliation{World Bank Group, Washington, DC, USA}
	\author{Bruno Lepri}
	\affiliation{Fondazione Bruno Kessler, Trento, Italy}
	\author{Nancy Lozano-Gracia}
	\affiliation{World Bank Group, Washington, DC, USA}
	\author{Samuel P. Fraiberger$^{ \dagger,}$ }
	\affiliation{World Bank Group, Washington, DC, USA}
	\affiliation{Massachusetts Institute of Technology, Cambridge, MA, USA}
	\affiliation{New York University, New York, NY}
	
	\def\thefootnote{$*$}\footnotetext{These authors contributed equally to this work}
	\def\thefootnote{$\dagger$}\footnotetext{To whom correspondence should be addressed: sfraiberger@worldbank.org, lorenzo.lucchini@unibocconi.it}
	
	\begin{abstract}
		Mobile phone data have played a key role in quantifying human mobility during the COVID-19 pandemic. Existing studies on mobility patterns have primarily focused on regional aggregates in high-income countries, obfuscating the accentuated impact of the pandemic on the most vulnerable populations. 
		Leveraging geolocation data from mobile-phone users and population census for 6 middle-income countries across 3 continents between March and December 2020, we uncovered common disparities in the behavioral response to the pandemic across socioeconomic groups. Users living in low-wealth neighborhoods were less likely to respond by self-isolating, relocating to rural areas, or refraining from commuting to work. The gap in the behavioral responses between socioeconomic groups persisted during the entire observation period. Among users living in low-wealth neighborhoods, those who commute to work in high-wealth neighborhoods pre-pandemic were particularly at risk of experiencing economic stress, facing both the reduction in economic activity in the high-wealth neighborhood and being more likely to be affected by public transport closures due to their longer commute distances. While confinement policies were predominantly country-wide, these results suggest that, when data to identify vulnerable individuals are not readily available, GPS-based analytics could help design targeted place-based policies to aid the most vulnerable.
	\end{abstract}

	\maketitle

	\section{Introduction}\label{sec:Introduction}
	Since its emergence in late 2019, SARS-CoV-2 has led to millions of infections and deaths worldwide, disrupted economies, strained healthcare systems, and caused social and psychological challenges for many individuals and communities at an unprecedented scale. With no vaccine in sight in the early stages of the pandemic, governments and local authorities quickly implemented non-pharmaceutical interventions (NPIs), such as stay-at-home orders or workplace closures, aiming to reduce physical contacts~\cite{tian2020investigation,gatto2020spread,chinazzi2020effect,kraemer2020effect,perra2021non,lucchini2021living}. Such measures coupled with the fear of the virus triggered an abrupt reduction in mobility, which contributed to slowing down the spread of the disease ~\cite{pullano2020population,woskie2021,lai2020effect,nouvellet2021}. However, these policies were predominantly untargeted, with little regard to socioeconomic status or need, prompting questions on the differentiated impact of the pandemic on the most vulnerable populations~\cite{checo2022assessing}.\\
	
	Over the past decade, mobile phone data have become the primary source of real-time disaggregated information on human movements~\cite{fudolig2021internal, alexander2015origin,jiang2017activity}, and this trend has accelerated during the pandemic~\cite{aleta2020modelling,yabe2022mobile,aleta2022quantifying,pangallo2022unequal}. Quantifying human mobility in real-time has been key to anticipating the evolution of the virus and the resulting economic shock. Previous works quantifying mobility during the pandemic, from epidemiological surveillance~\cite{jia2020population,grantz2020use} to policy impact evaluation~\cite{bonaccorsi2020}, predominantly focused on high-income countries~\cite{badr2020association,Weill19658,lucchini2021living,kraemer2021spatiotemporal}. While some of these studies have shown that the pandemic has had a more pronounced impact on the most vulnerable~\cite{paul2021socio,jay2020neighbourhood,gauvin2021,blundell2020covid,bambra2020covid,dorn2020,abedi2021} and that, in general, different mobility patterns are connected with income segregation~\cite{moro2021mobility}, little is known on the socioeconomic disparities in mobility behavior across middle-income economies. Existing studies on mobility in middle-income countries either focused on aggregated trends without taking into account socioeconomic backgrounds~\cite{maloney2020determinants, kephart2021effect} or on country-specific case studies~\cite{gozzi2021estimating,heroy2021covid,mena2021socioeconomic}.
	
	Here, we provided a fine-grained analysis of human mobility across 6 middle-income countries, spanning 3 continents, during the period from March 2020 to December 2020. Using GPS location data from the mobile device applications of 281 million users, we employed state-of-the-art methodologies based on both spatial and temporal clustering to accurately infer how a user allocate their time between their home, their workplaces, and other locations that they visit (see Sec. SI~3A and 3B)~\cite{lucchini2021living}. In the absence of income or consumption data, we assign each user an asset-based wealth proxy derived from census data on the administrative unit where they live, we then characterized the propensity of mobile phone users of various socioeconomic groups to self-isolate at home, to relocate to a rural area, or to commute to work.
	
	We found that the wealth of the neighborhood where a user lives is a strong determinant of their mobility behavior during the pandemic, and that mobility gaps between users living in neighborhoods with different levels of wealth are remarkably consistent across countries (although showing different amplitudes), supporting previous findings on single countries~\cite{gozzi2021estimating,heroy2021covid,mena2021socioeconomic}. Relative to the pre-pandemic period, users living in a high-wealth administrative unit of a metropolitan area (defined as the top wealth-ranked areas, where 20\% of the population lives) had a self-isolation rate 111 percentage points higher than those living in the bottom 40\% (``low-wealth place''), their rate of relocating to rural areas once the pandemic hit was 49 percentage point higher, and their reduction in commuting to work was also higher by 30 percentage points. While users' mobility slowly started to revert toward pre-pandemic levels, the gap in the behavioral response across socioeconomic groups persisted over the entire observation period. Furthermore, by specifically focusing on users living in low-wealth places, we found that those who used to commute to high-wealth places prior to the pandemic stopped commuting 40\% more during the observation period than those who used to commute to low-wealth places. Using a dataset of policy actions standardized across countries, we also discovered that the closure of public transportation was associated with a stronger reduction in commuting for users living in low-wealth neighborhoods who used to commute to high-wealth places, whereas we do not find a significant association for those commuting to low-wealth places. As the ability of the poor to work from home is extremely limited in most developing countries and in particular for low-wealth individuals~\cite{garrote2021earth,dingel2020many}, these findings suggest that attention to vulnerable groups is needed when untargeted policies are implemented if aiming to reduce the potential economic stress individuals might face. 
	
	Our results also indicate that, when data to identify vulnerable individuals are not available, mobile phone data can provide useful information to identify key changes in population behavior and contribute to better-informed policy-making, through improved targeting of vulnerable groups. While our analysis cannot exclude alternative explanations for these findings, our approach can work in support of traditional methods providing additional guidance for practitioners and policymakers.
	
	\section{Results}\label{sec:Results}
	\begin{figure*}[!ht]
		\centering
		\includegraphics[width=1\textwidth]{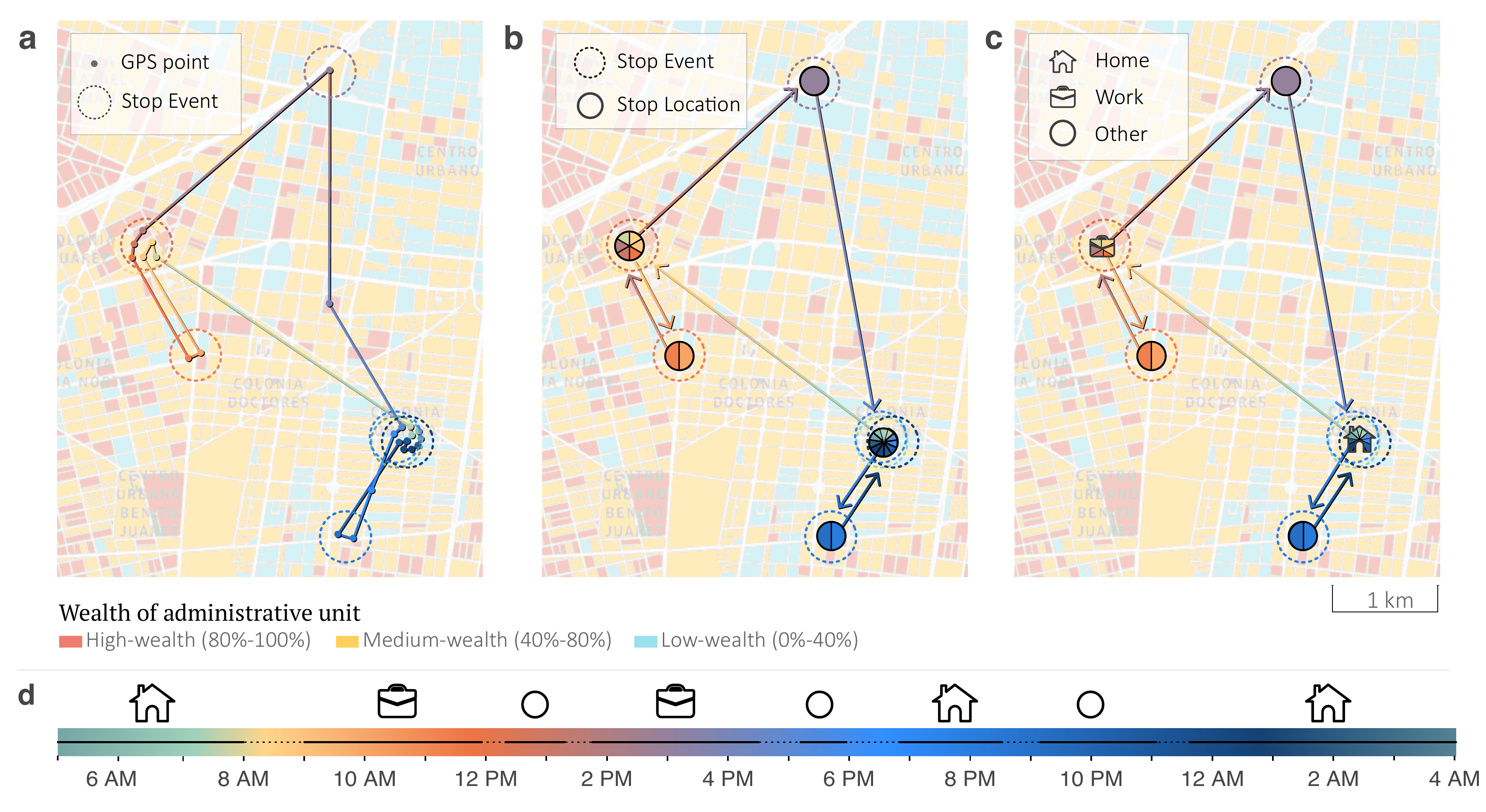}
		\caption{\emph{Inferring location type and time use from GPS trajectories.} \textbf{(a)} Trajectory of a hypothetical mobile phone user over one day. A stop event (dotted circle) characterizes a location where a user spends at least 5 minutes within a 25-meter distance. The color of a stop event corresponds to the time of the day when the event occurs, and lines connect two consecutive stop events. \textbf{(b)} Stop events are spatially clustered together to form a stop location. \textbf{(c)} Stop locations are then labeled as \textit{home}, \textit{workplace}, or \textit{other} based on how frequently they are visited during the observation period and the hours of the day at which a visit occurs. \textbf{(d)} Final structure of the user location data, which consists of a non-continuous time series of labeled stop events. The background image shows a portion of Mexico City where blocks are colored by the level of a wealth index constructed from census data, illustrating the granularity of our data in urban areas.}
		\label{fig:drawingProcessing}
	\end{figure*}
	
	\subsection{Inferring location type and time use}
	Our GPS dataset comes from Veraset~\cite{VerasetMovementVeraset} and contains the anonymized timestamped geocoordinates of 281 million mobile phone users located in 6 middle-income countries --Brazil, Colombia, Indonesia, Mexico, Philippines, and South Africa-- between January 1st and December 31st, 2020 (``observation period''). These countries were selected to cover diverse regions of the globe to make the results as general as possible from a geographical, cultural, and policy-implementation perspective (see Sec. SI 12A)~\cite{hale2020oxford}. In this study, we focus on the longitudinal behavior of users active on at least $20\%$ of days during the observation period and at least $20\%$ of days between January 1st and March 15th (``pre-pandemic period''). This, on average, covers about 1.35\% of each country's urban population. We also refer to the period from February 1st to March 15th $2020$ as the ``baseline period'' as it will serve as a comparison unit of individual behavior from before to after the pandemic declaration (see Figures SI 15-18 for more details on changes in mobility indicators during the baseline period).
	
	To uncover how users spent their time, we first applied a spatiotemporal clustering algorithm to convert their GPS coordinates into a sequence of stop events where a user spent at least 5 minutes within a 25-meter distance (Figure~\ref{fig:drawingProcessing}). We then spatially clustered stop events to identify unique locations repeatedly visited by a user over time (``stop locations''). Finally, we classified stop locations as \textit{home}, \textit{workplace}, or \textit{other} based on how frequently they were visited during the observation period and the hours of the day at which a visit occurred. More specifically, home and work locations were computed over a window of $49$ days, requiring them to be visited at least on $20\%$ of the days on which a user was active within the window. Additionally, work locations were required to be visited for at least $1$ hour per day on average when visited. Home and work location inference was based only on visit patterns during respectively nighttime (from Monday to Friday, from $11$ P.M. to $5$ A.M. of the following day) or weekends (Saturday and Sunday), and working days daytime (from Monday to Friday, from $5$ A.M. to $11$ P.M.). A sample of 500 users' stop locations' sequences (with their respective duration and time of visit) were manually annotated by two independent individuals to provide a supervised set of labels. This enabled us to obtain optimal parameter values for our classifier of location type, which reached an average agreement of $80\%\pm 3\%$ with the manual home-work label assessment (see Sec. SI~5 for more details on performances and errors' estimates). 
	Taken together, these steps allowed us to convert a mobile user trajectory into a non-continuous time series of stop events labeled by location type.
	
	\subsection{Classifying users by wealth}
	In the absence of income or consumption data for individual users, we used the most recent population census in each country to generate an asset-based proxy of their wealth from that of the administrative unit where they live~\cite{mckenzie2005measuring} (see Table~\ref{tab:CountryUsersEventsClusters} for country specific median areas of administrative units). We estimated a one-dimensional wealth index from data on asset ownership and access to services for the most disaggregated level of administrative units in each country (see Materials and Methods). We then assigned to each user the wealth index of the administrative unit where their most frequently visited home during the pre-pandemic period was located (``primary home''). This procedure allowed us to quantify how users who differ by the wealth of their primary home's administrative unit allocated their time between their homes, their workplaces, and other locations over time. Based on the wealth associated with each of these locations we will refer to users living or working in a specific location as being part of the corresponding ``wealth group''. The association between a user living or working in a specific location and their wealth cannot be precisely estimated from the data at our disposal and should thus be only intended as a proxy of such measure. In what follows, among all available active users in our dataset, we restricted our analysis to the $46\%$ of them whose primary home is located in one of the five most populated cities of each country, where mobile phone users are concentrated (see Tab.~\ref{tab:CountryUsersEventsClusters} in Materials and Methods, and SI Sec. 13). 
	
	\subsection{Self-isolating at home} \label{sec:wg_home}
	
	\begin{figure*}[!ht]
		\centering
		\includegraphics[width=1\textwidth]{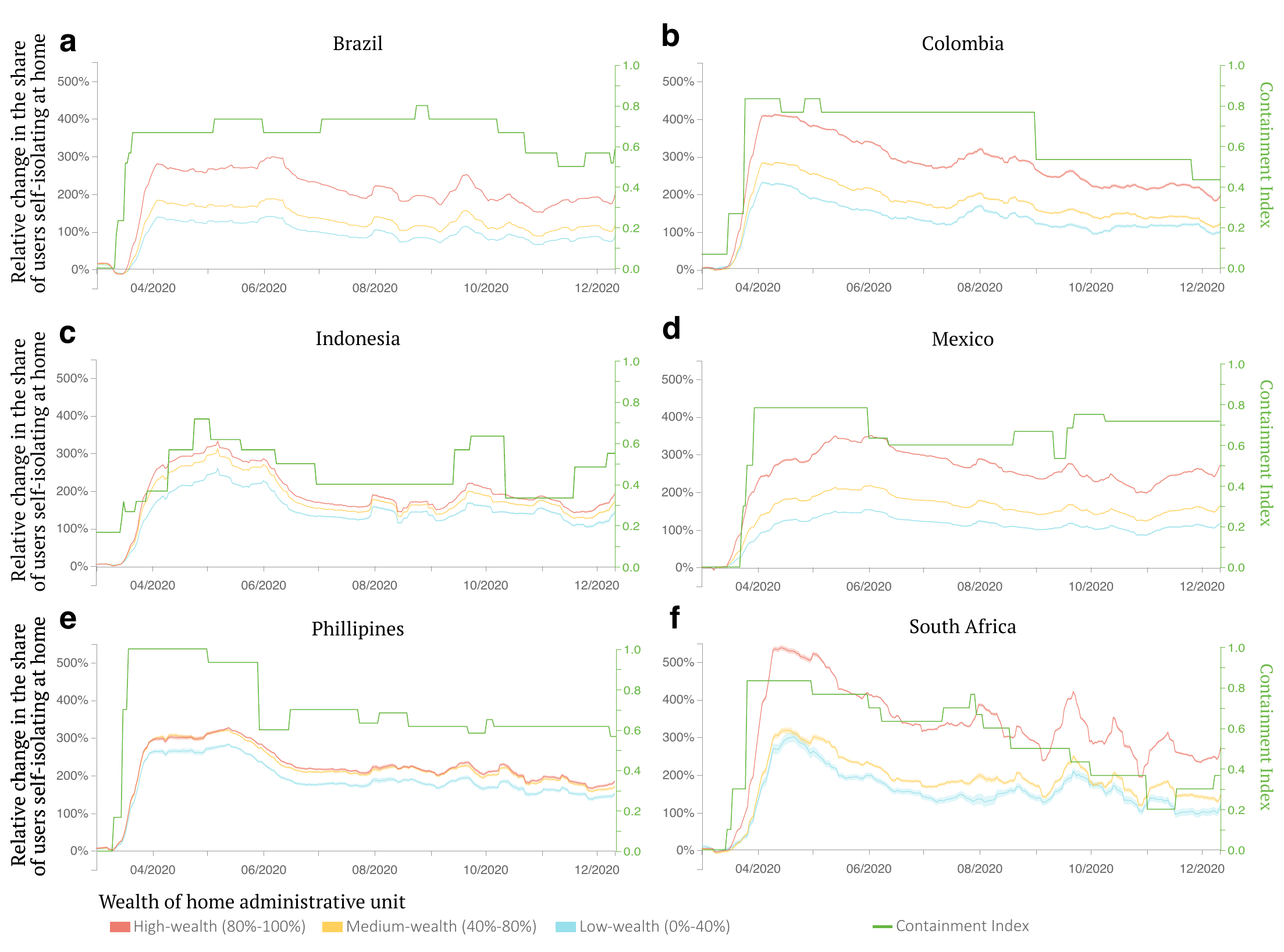}
		\caption{\emph{Change in the share of users self-isolating at home by socioeconomic group.} Each panel shows the relative change in the share of active users staying at home over the entire course of a day relative to the pre-pandemic period for the six countries studied, conditioning on the wealth of their primary home administrative unit. Shaded areas indicate the standard errors of the mean computed pulling together the self-isolating share of all administrative units of a single country. In countries where many administrative units are home to users, the standard error becomes very small due to the higher number of geographical areas included in the computation.} We also report the stringency index of containment policies in each country over time (green line). The values of the stringency index of containment policies are reported on the y-axis on the right of each panel. Across all countries, users living in high-wealth places were more likely to isolate at home when the pandemic hit than those living in low-wealth places, and the gap persisted during the observation period.
		\label{fig:home_mob}
	\end{figure*}
	
	To characterize mobility behavior during the pandemic, we started by focusing on the propensity of users to self-isolate at home relative to the pre-pandemic period (Fig. ~\ref{fig:home_mob}). In the early stages of the pandemic when no vaccines were available, home self-isolation was one of the solutions to which most of the countries resorted for rapidly controlling the infections~\cite{hale2020oxford}. Formally, we define a  mobile phone user to be ``self-isolating'' at home on a specific day if they are visiting exclusively their home location on that specific day. We find that, on average, the share of users living in high-wealth neighborhoods and self-isolating at home was 252\% higher than the baseline period, compared to a 141\% increase in users living in low-wealth neighborhoods, a 111 percentage points difference. The gap between high- and low-wealth groups' propensity to self-isolate is observed across all 6 countries in our sample, ranging from a difference of 36 percentage points for the Philippines to 175 for South Africa. While 55\% of users kept spending some of their time outside of home, time spent at home increased on average by 19\% for the high-wealth group versus 13\% for the low-wealth group. The reduction in time spent outside of the home was primarily driven by a reduction in time spent at work, which dropped by 40\% on average for the high-wealth group while only 29\% for the low-wealth one after the pandemic declaration (Fig. SI~3), suggesting that the capacity to work from home could be one of the main determining factor{s in the decision to self-isolate~\cite{garrote2021earth}. Although the mobility of all socioeconomic groups gradually started to revert back to its pre-pandemic level, the gap between high- and low-wealth groups persisted, indicating that low-wealth groups remained more exposed to physical contact during the observation period.
		
		\subsection{Relocating to rural areas}
		
		\begin{figure*}[!ht]
			\centering
			\includegraphics[width=1\textwidth]{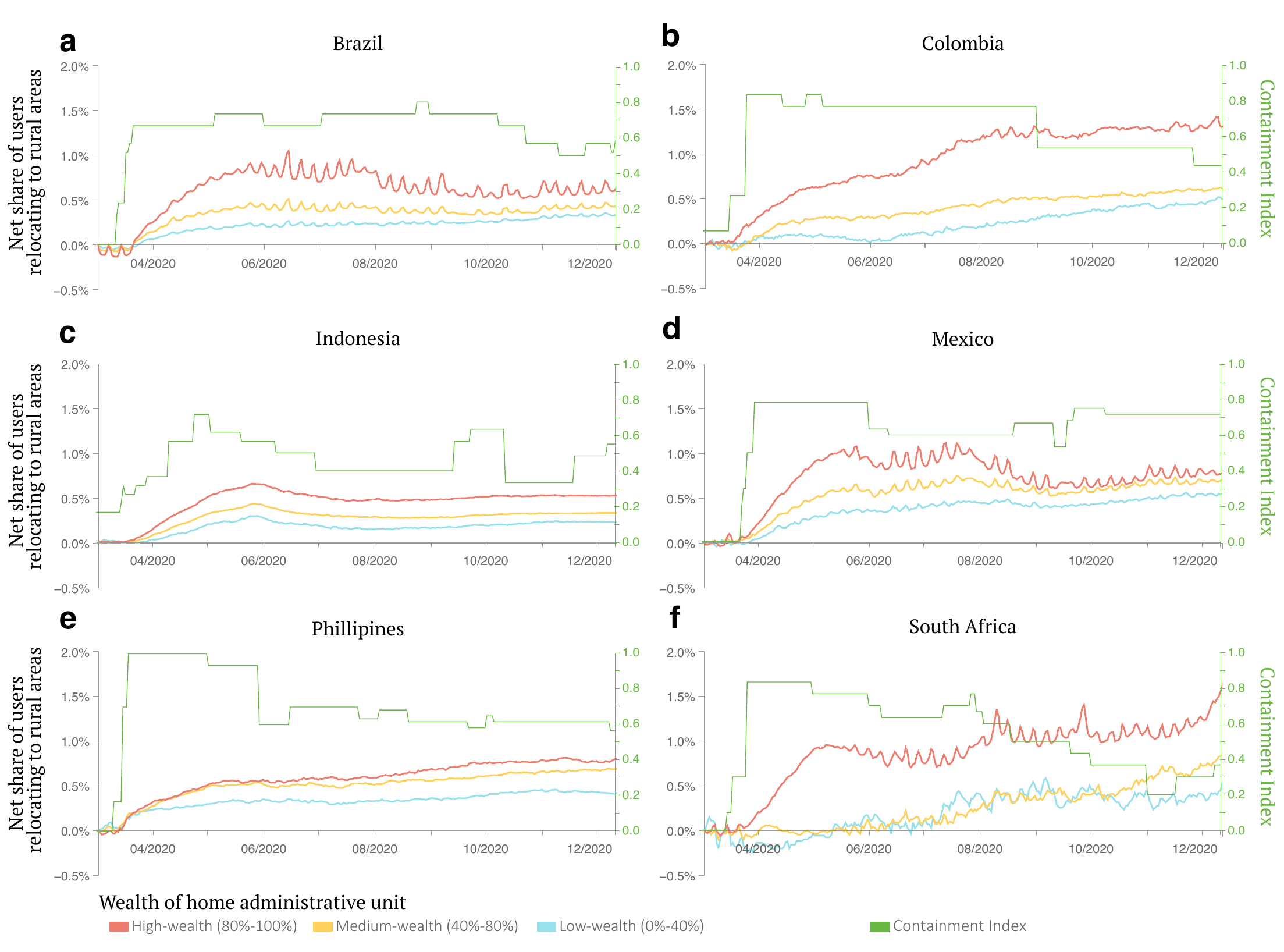}
			\caption{\emph{Net share of urban users relocating to rural areas by socioeconomic group.} Each panel shows the percentage change in the difference between the number of users relocating from an urban to a rural area and those moving from rural to urban, for different socioeconomic groups in the six countries under study. Results are normalized to remove pre-pandemic relocation flows. We also report the stringency index of containment policies in each country over time (green line). The values of the stringency index of containment policies are reported on the y-axis on the right of each panel. Across all countries, users whose primary home is located in a high-wealth neighborhood (red line) were more likely to relocate to a rural area than those living in a low-wealth neighborhood (blue line).}
			\label{fig:home_mig}
		\end{figure*}
		
		Quantifying population movements between urban and rural areas is key to uncovering how the virus propagated geographically within countries and provides some insights into the capacity of mobile phone users to respond to the evolution of the pandemic and mobility restrictions.
		Thanks to our dynamic classification of home locations, we could identify users relocating during the observation period (Fig. \ref{fig:home_mig}). Across all 30 cities in our sample, we find that a net flow of about 0.61\% of users relocated to rural areas during the first 3 months of the pandemic. users living in high-wealth neighborhoods were proportionally more likely to relocate to rural areas compared to those living in low-wealth neighborhoods, with an average difference of 49 percentage points between these two groups. The average gaps vary across countries, from a 27 percentage points difference for the Philippines to an 80 percentage points difference for South Africa, which faced the largest relocation gap between socioeconomic groups. Relocation flows then remained relatively flat in the latter half of 2020. These patterns can be explained by different non-exclusive dynamics. For example, they could reflect differences between users from different wealth groups in consumed information or in information source preferences. Moreover, these are also compatible with the more risk-averse behavior of users living in high-wealth neighborhoods having more options than users living in low-wealth neighborhoods to relocate to less densely populated rural areas to minimize physical contact~\cite{coven2022jue}. Similar results were obtained in high-income countries, providing evidence of urban flight of predominantly wealthier populations. For the case of the US, this is further linked to the potential acceleration of the virus propagation caused by users living in high-wealth neighborhoods having relocated to rural areas~\cite{coven2022jue,EscapefromNewYork}. While our data cannot establish a direct causal link in support of this interpretation, these results signal an important aspect for policymakers to pay attention to.
		
		\subsection{Commuting patterns}
		
		\begin{figure*}[!ht]
			\centering
			\includegraphics[width=1\textwidth]{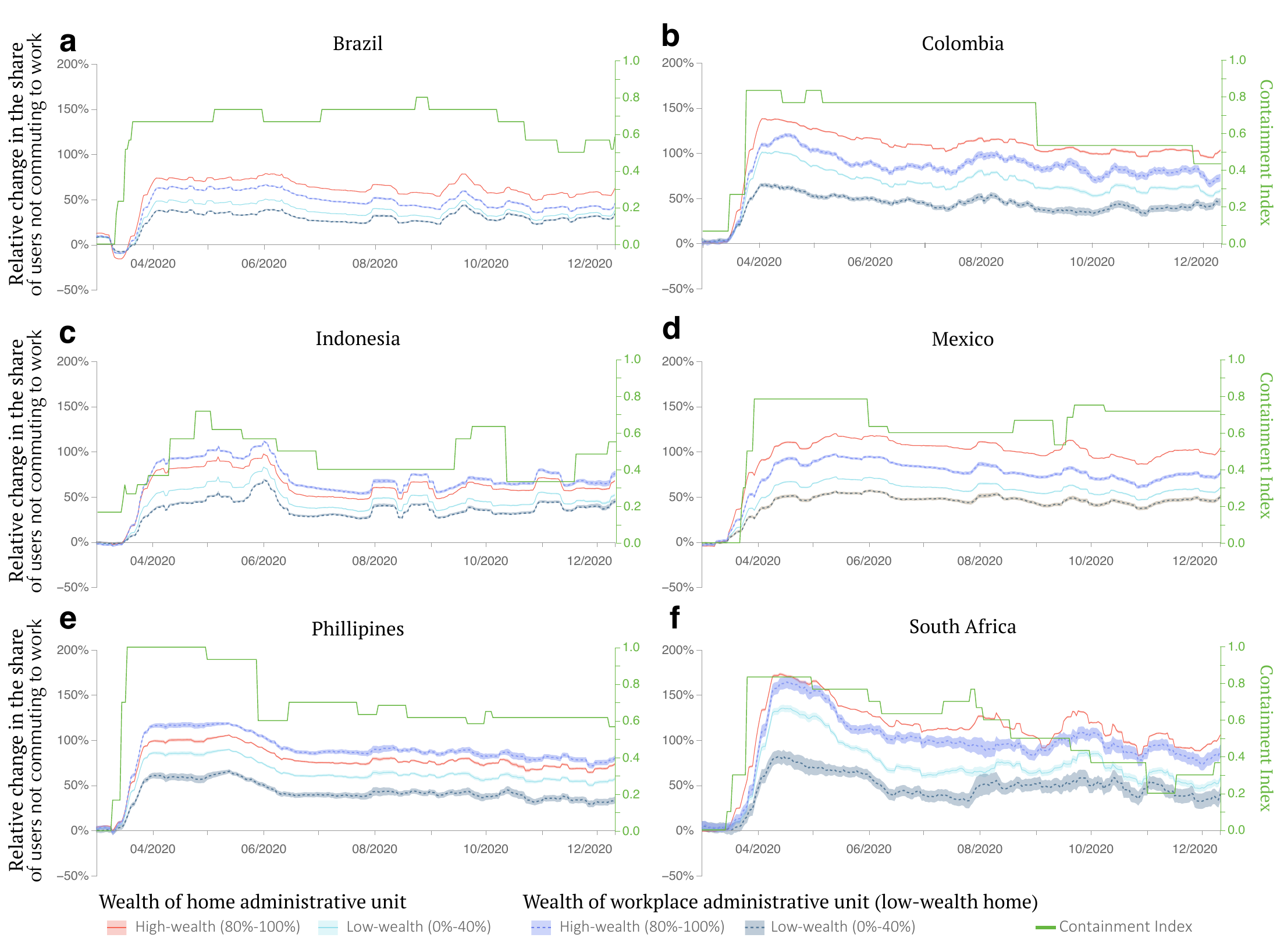}
			\caption{\emph{Change in the fraction of users not commuting by socioeconomic group.} Focusing on the 26\% of users with a work location during the observation period, we illustrate the percentage change in the number of users who are not commuting, conditioning on their wealth group classification. Users in the high-wealth group (solid red line) were more likely to stop commuting than those in the low-wealth group (solid light-blue line). We then restrict to users from the low-wealth group and measure their changes in commuting patterns conditioning on the wealth of their workplace. Users living in low-wealth neighborhoods who used to work in high-wealth neighborhoods pre-pandemic (dashed violet line) were more likely to stop commuting than those who used to work in low-wealth neighborhoods (dashed grey-blue line). The shaded region highlights the standard errors of the mean computed pulling together the share of users not commuting to their workplace from all administrative units of a single country. The green line shows the stringency of containment policies in the corresponding countries over time. The values of the stringency index of containment policies are reported on the y-axis on the right of each panel.}
			\label{fig:work_mob}
		\end{figure*}
		
		A key challenge of untargeted confinement policies stems from the uneven ability to work from home across socioeconomic groups. While this issue has extensively been documented in high-income countries, it is exacerbated in developing countries where only a small fraction of rich individuals can work remotely due to the structure of labor markets, higher levels of informal employment, reduced internet access, and a lack of compensatory income support~\cite{garrote2021earth}. To shed light on these issues, we estimate users' propensity to commute relative to the pre-pandemic period (Fig. ~\ref{fig:work_mob}). Although we do not have fine-grained measures of changes in unemployment, under the assumption that users living in low-wealth neighborhoods lack the ability to systematically work from home~\cite{garrote2021earth}, a change in commuting behavior is a good proxy for a change in employment status for this group. Formally, we define a user to be ``commuting'' on a specific day if they are visiting their work location on that specific day. We find that the share of users who stopped commuting increased sharply in all countries in the early stage of the pandemic. It then reverted before stabilizing in the latter half of 2020. 
		Consistent with our findings on self-isolating behavior, the share of users living in high-wealth neighborhoods who stopped commuting once the pandemic started was on average 30 percentage points higher compared to users living in low-wealth neighborhoods. The difference between high- and low-wealth groups ranges from 15 percentage points for the Philippines to 43 percentage points for Mexico. We then separated users living in low-wealth neighborhoods by the wealth of the administrative unit where they used to commute during the pre-pandemic period. Commuters to high-wealth neighborhoods had a larger share of users who stopped commuting by 37 percentage points compared to commuters to low-wealth neighborhoods. These results are consistent with existing studies of high-income countries~\cite{chetty2020economic, mongey2021workers} that connect the decision of users living in high-wealth neighborhoods to self-isolate with the reduction in demand for goods and services in high-wealth neighborhoods where they live. Findings for high-income countries report that this dynamic, in turn, predominantly impacted the employment prospects of users living in low-wealth neighborhoods who used to work in high-wealth neighborhoods before the pandemic started~\cite{chetty2020economic, mongey2021workers}.
		
		\subsection{Policy restrictions and mobility behavior}
		\label{sec:modeling_policy}
		
		\begin{figure*}[ht]
			\centering
			\includegraphics[width=.7\linewidth]{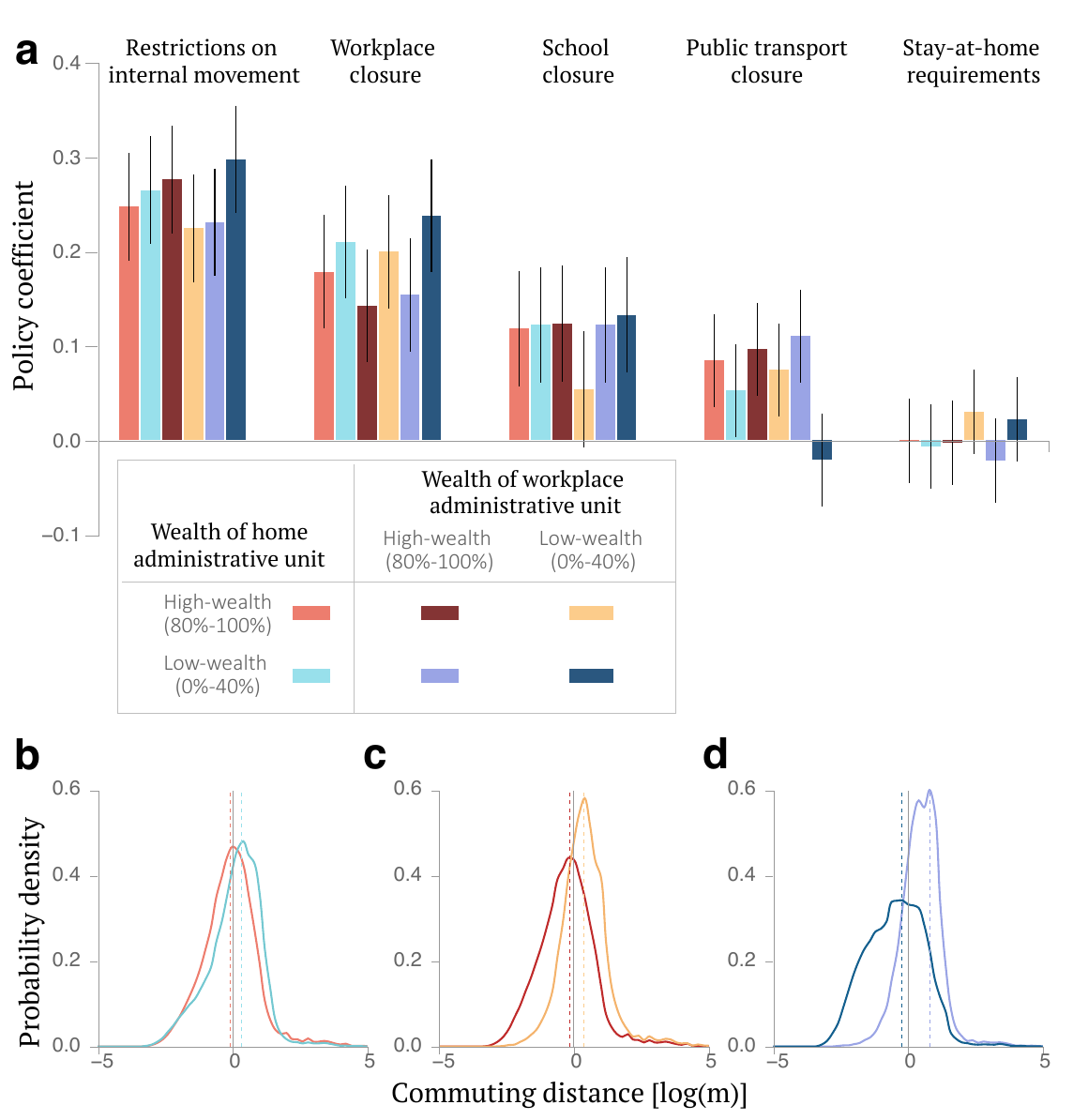}
			\caption{\emph{Policy restrictions and commuting patterns across socioeconomic groups.} 
				a) We present the estimated coefficients for policy restrictions, $c_{i,n}$, modeling the propensity of users from different socioeconomic groups to suspend commuting. Errorbars report the $95\%$ confidence intervals for the estimated coefficient values. b-d) Distributions of the distance between home and workplace across socioeconomic groups, b) comparing users living in low- versus high-wealth neighborhoods, then c) focusing on users living in high-wealth neighborhoods and comparing those working in low- versus high-wealth neighborhoods, and d) focusing on users living in low-wealth neighborhoods and comparing those working in low- versus high-wealth neighborhoods.}
			\label{fig:conditioned_model_parameters}
		\end{figure*}
		
		Taken together, our results indicate that users' mobility changed sharply during the first weeks of the pandemic, predominantly for users located in high-wealth neighborhoods, generating a socioeconomic gap that persisted during the entire observation period. These findings call into question the extent to which government interventions imposing restrictions on mobility to reduce the spread of infections contributed to this gap. We, therefore, analyzed the statistical associations between containment measures --restrictions on internal movement, school closure, workplace closure, public transport closure, and stay-at-home requirements-- and mobility. We specifically focused on the fraction of users commuting to work across wealth groups, an outcome of utmost importance to characterize the distributional association between containment policies and jobs in a timely fashion. We estimated a multivariate panel regression model for each wealth group, which includes local and global incidence of cases obtained from Our World in Data~\cite{covid19datapublicdata}, and stringency indices of policy restrictions from the Oxford COVID-19 Government Response Tracker~\cite{hale2020oxford}. All the explanatory variables are independently standardized for each wealth group, therefore our panel regression with fixed effects can be used to elicit relative differences between group-specific responses to containment measures (see Materials and Methods).
		
		The majority of policy restrictions during the observation period were implemented country-wide within a short time frame in the first weeks of the pandemic when people's risk perception of the virus also shifted abruptly (Fig. SI~31). It is therefore difficult to tease out the effect of a single policy enactment happening during this period. However, while in the immediacy of an aggravating emergency most of the policies were simultaneously enacted, after this first period (roughly corresponding to the first month since the pandemic declaration; see Fig. SI 31) enactments and reopenings became less synchronous and more diverse. For this reason and to provide a more precise estimate of the association between policy changes and mobility behavior changes, in this section, we focus on the period from April $11$, $2020$ --disregarding the first month after the pandemic declaration-- to December $31$, $2020$ (see Sec. SI~11B for sensitivity analysis of the results). 
		Except for stay-at-home requirements, we found that all containment policies were associated with a higher fraction of users who stopped commuting to work (Fig~\ref{fig:conditioned_model_parameters}a). As we excluded the first month of the pandemic from the analysis, we did not find a significant difference in mobility response between users living in high- and low-wealth neighborhoods. However, by focusing specifically on users living in low-wealth neighborhoods, we found that public transport closures were associated with a significant reduction in commuting for users commuting to high-wealth neighborhoods (coefficient=$0.10$, $95\%$ confidence interval$=[0.05,0.15]$). By contrast, we were not able to find a significant association with the implementation of such measures for users living in low-wealth neighborhoods.
		These findings reflect the fact that users living in low-wealth neighborhoods and working in high-wealth neighborhoods are those commuting the longest distances on average, therefore often relying on public transport to get to work, while those working in low-wealth neighborhoods have the shortest commuting distances (Fig. \ref{fig:conditioned_model_parameters}b-d). Without conditioning by the wealth of workplace neighborhood, average commuting distances are not statistically different between wealth groups based on their home location, stressing the value of having detailed information on individual users to characterize their behavior. Case incidence was also found to be a significant determinant of self-isolation and migration for users living in high-wealth neighborhoods, as found in previous work focusing on the early period of the pandemic ~\cite{maloney2020determinants}.

		\section{Discussion}\label{sec:Discussion}
		
		GPS data from personal mobile devices have played a key role in providing timely information to quantify the impact of the COVID-19 pandemic. Here, we focused on six middle-income countries~\cite{WB_class}--Brazil, Colombia, Indonesia, Mexico, Philippines, and South Africa-- from Latin America, Sub-Saharan Africa, and South East Asia to elicit common disparities across socioeconomic groups in the behavioral responses to the pandemic and containment measures.
		By analyzing longitudinal data over one year, we quantified the behavior of mobile phone users living in urban areas in terms of their propensity to self-isolate at home, relocate to rural areas, and suspend their daily commute. 
		The task is particularly challenging as GPS data coverage in middle and low-income countries is known to differ significantly from high-income countries~\cite{wang2024infrequent, yang2023identifying, edsberg2022understanding, heroy2021covid}. In our case, while aggregating thousands of SDK providers helps reduce the gap, we find significant differences in data coverage across different countries.
		
		While these differences make it impossible to generalize the claims to the general urban population of these six countries, we nevertheless uncovered a consistent socioeconomic gap in the percentage of users who adapted their mobility behavior in response to the pandemic. users from high-wealth socioeconomic groups were more likely to self-isolate at home and to relocate to rural areas to reduce their exposure to the virus. A greater percentage of users living in high-wealth neighborhoods stopped their daily commute, reflecting their higher propensity to work from home. On average, we observed wider gaps between high and low-wealth groups in upper-middle-income countries (Brazil, Colombia, Mexico, and South Africa~\cite{WB_class}) than in lower-middle-income countries (Indonesia and the Philippines~\cite{WB_class}). This potentially indicates a greater ability of users living in high-wealth neighborhoods in these countries to work from home than for lower-middle-income countries ~\cite{garrote2021earth}. 
		
		When we focused specifically on the commuting patterns of users living in low-wealth neighborhoods --who, to a higher degree, lack the ability to work from home-- we discovered that in all six countries, those commuting to high-wealth neighborhoods pre-pandemic were more likely to stop commuting during the pandemic than those commuting to low-wealth neighborhoods. For example, by limiting their outside activities or abandoning urban areas, users living in high-wealth neighborhoods might have amplified the economic shock affecting, in turn, also users living in low-wealth neighborhoods working in high-wealth neighborhoods. Furthermore, owing to their longer commuting distances, users living in low-wealth neighborhoods and working in high-wealth neighborhoods were also more likely to stop commuting following public transport closures, whereas the commuting behavior of those working in low-wealth neighborhoods was not found to be significantly affected by this policy restriction. Taken together, these findings indicate that mobile phone users living in low-wealth neighborhoods and working in high-wealth neighborhoods were disproportionately burdened in their ability to work. 
		
		These results illustrate the delicate balance between ensuring a forceful response to a pandemic and the unintended consequences resulting from untargeted interventions, which could disproportionately affect economically vulnerable groups. While targeting based on individual information is always preferred, information may not be available or it may be difficult to acquire in cases of emergency where time is of the essence. In such cases, place-based policy interventions may provide an alternative for effective targeting and optimally distributing additional support to the most vulnerable. As developing countries often lack the possibility to access up-to-date information on individuals~\cite{world2021world}, mobile data could provide a tool to implement such targeted policies in a timely fashion and respond to future pandemics more appropriately.

		\section{Materials and Methods}\label{sec:Methods}
		\subsection{Datasets}
		
		\subsubsection*{Population census}\label{sec:data-wealth}
		We collected data from the most recent population census in each country to construct a wealth index measuring the socioeconomic level of an administrative unit. We use data on the level of education, access to health services, assets' ownership, and various household and social characteristics for the smallest administrative unit available. These different dimensions are aggregated into a single index using principal component analysis to produce a one-dimensional wealth index of an administrative unit following the procedure described in ~\cite{fraiberger2020uncovering, vyass2006constructing}.
		
		\subsubsection*{Administrative boundaries}\label{sec:data-admin}
		Administrative boundaries consist of geolocalized polygons with unique identifiers. They are retrieved from the National Institute of Statistics or the National Institute of Geography website for each country. The spatial resolution level varies depending on the country. All the administrative boundary files share at least four common administrative levels: national, regional, urban, and suburban units. Administrative boundaries are used to compute the wealth index and mobility patterns for different socioeconomic groups.
		
		\subsubsection*{Urban extents}\label{sec:data-urban}
		Urban extents are obtained from the GHS Urban Centre Database (GHS-UCDB). It characterizes spatial entities called “urban centers” according to a set of multitemporal thematic attributes gathered from the Global Human Settlement Layer sources and integrated with other sources available in the open scientific domain. The urban centers are defined by specific cut-off values on resident population and built-up surface share in a 1x1 km uniform global grid. As such, urban extents are defined as contiguous cells (without diagonals and with gap filling) with a density of at least 1,500 inhabitants/$km^2$ and a minimum of 50,000 inhabitants. This dataset has global coverage and is therefore well-suited for multi-country applications. 
		
		\subsubsection*{Veraset movements}\label{sec:data-veraset}
		Mobile data are provided by the data company \textit{Veraset}. It consists of the anonymized timestamped GPS coordinates of mobile devices, covering about $5\%$ of the global population. The data is sourced from thousands of Software Development Kits (SDKs), potentially helping reduce sampling biases that might arise from gathering data only from a limited number of apps or from apps exclusively dedicated to specific tasks~\cite{VerasetMovementVeraset}.
		
		\subsection{Mobility data processing}\label{sec:mobilityDataProcessing}
		We present a brief description of the procedure to infer relevant information about individual users. We refer to a single GPS point within a user's trajectory as ``ping''. Each ping has an associated accuracy measure providing an estimate of the coordinates' precision. Starting from a collection of trajectories we perform a series of steps to identify the types of locations being visited by a user \cite{lucchini2021living}: 
		
		\begin{enumerate}
			\item First, we aggregate pings into stop events, which are clusters of spatiotemporally contiguous pings (for more details, see Sec. SI~3A).
			\item Second, we perform an additional spatial clustering step to aggregate stop events that are close enough to be associated with a single location.
			\item Third, to reliably assign a user home location, and to subsequently connect its demographic information, we follow a consolidated approach, which makes use of different heuristics based on circadian rhythms and weekdays-weekend patterns~\cite{csaji2013exploring,ccolak2015analyzing,alexander2015origin,pappalardo2020individual,song2010modelling,barbosa2015effect}(see Sec. SI~3C for details).
			\item Fourth, we restrict the set of users to those who appear in our records at least once per day for at least $20\%$ of the days pre-pandemic, and $20\%$ of the days during the observation period (see Sec. SI~3B for additional details). 
		\end{enumerate}
		
		After these processing steps, we associate each user with a wealth proxy based on the administrative unit where their primary home location is located (Tab.~\ref{tab:CountryUsersEventsClusters}). To precisely connect mobility with demographic data at the smallest sub-urban scale available, further processing is required, which we describe in Sec. SI~4.

		\begin{table}[!ht]
			\centering
			\setlength{\tabcolsep}{5.pt}
			\begin{tabularx}{0.8\columnwidth}{@{}X rrrrrr@{}}
				\toprule
				& \textbf{BR} & \textbf{CO} & \textbf{ID} & \textbf{MX} & \textbf{PH} & \textbf{ZA}\\
				\midrule
				\textbf{Active users} & 1514679 & 142933 & 1205827 & 676775 & 104681 & 144445 \\
				\textbf{GPS points} & 3.9e10 & 2.9e9 & 1.9e10 & 1.7e10 & 1.9e9 & 3.4e9\\
				\textbf{Stop events} & 1452M & 103M & 893M & 656M & 75M & 154M \\ 
				\textbf{Stop locations} & 111M & 8M & 76M & 50M & 5M & 12M \\
				\textbf{Urban Pop.} & 1.7\% & 0.5\% & 1.0\% & 1.6\% & 0.2\% & 0.7\%\\
				\textbf{Total Pop.} & 0.7\% & 0.3\% & 0.4\% & 0.5\% & 0.09\% & 0.2\%\\
				\midrule
				\textbf{Admin }\small{\textit{covered}} & 100.00\% & 32.18\% & 97.96\% & 41.01\% & 93.13\% & 80.00\%\\
				\textbf{Admin }\small{\textit{users}} & 2.37\% & 1.15\% & 0.49\% & 1.41\% & 0.21\% & 0.92\%\\
				\textbf{Admin }\small{\textit{pop.}} & 16855 & 179 & 8497 & 85 & 3516 & 2739\\
				\textbf{Admin }\small{\textit{area}} & 3.052 & 0.006 & 3.449 & 0.008 & 0.200 & 0.983\\
				\textbf{Admin }\small{\textit{pop. dens.}} & 5750.98 & 32612.53 & 2306.29 & 11334.29 & 20258.07 & 2617.16\\
				
				\bottomrule
			\end{tabularx}
			
			\caption{Summary statistics of the dataset used for the analysis. Each column reports information for a different country, respectively, Brazil (BR), Colombia (CO), Indonesia (ID), Mexico (MX), the Philippines (PH), and South Africa (ZA). All numbers are for active users living in urban administrative units. Population in the ``Urban pop.'' and the ``Total pop'' rows report the fractions of active users against the urban and national population respectively. These were computed using census data from each country and taking into account only the population of the same urban administrative areas where the subjects had their primary home location. The ``Admin'' rows report respectively the percentage of administrative units in which at least one active user has their primary home location against the total number of urban administrative units (\textit{Admin covered}), the median population in the latest census available at the time of the study (\textit{Admin pop.}), the median coverage of users by the population in each administrative unit (\textit{Admin users}), the median size (reported in $km^2$) of administrative units used to associate a wealth category to active users living in urban areas (\textit{Admin area}), and the median population density of the given administrative units (\textit{Admin pop. dens.}), measured in $pop/km^2$ (see SI Sec. 13 for more details on the relationship between administrative units and the active user population).
			}
			\label{tab:CountryUsersEventsClusters}
		\end{table}
		
		\subsubsection*{Reweighting and wealth labels} 
		GPS data, sourced from mobile phones, are known to potentially introduce biases due to an uneven distribution of wealth among the devices' owners~\cite{oliver2020mobile,wesolowski2016connecting}. These biases need to be taken into account in the process of assigning labels to users and aggregating them into groups. In general, official demographic data are associated with single individuals based on the smallest sub-urban areas available for each country. While demographic data from the National Institutes of Statistics of each country (see Sec.~\ref{sec:data-urban}) can be considered as an unbiased information source, the process with which we link them with individual GPS information and aggregate them in groups afterward could drastically impact the reliability of the analyses. A simplistic approach would be to group users with lower wealth values, based on the wealth of the administrative unit their home location falls in. This would create groups based on mobile phone users' percentiles of wealth. In this framework, given a sample of $100$ users, the $40$ users with lower wealth would be labeled as ``{users living in low-wealth neighborhoods''. However, this would automatically transfer biases within the user base into biased wealth groups. Our population-based reweighting approach, in contrast, consists of a top-down classification of wealth groups. Administrative units with which at least one user is associated are considered and divided into wealth categories based both on their average wealth and on the fraction of the population living there. The population, as provided by local statistical authorities, of each administrative unit thus acts as a weight to reconstruct a group-specific representative population of each metropolitan area independently. High-wealth, Medium-wealth, and Low-wealth labels are then associated with each administrative unit considering discrete percentile groups. Thus, ``{users living in low-wealth neighborhoods'' will only be represented by those users living in administrative units whose wealth is in the lower $40\%$ of the wealth-index values for the population of a certain metropolitan area. We stress that we consider as a population the total number of residents living in all retained administrative units, divided by metropolitan areas, as provided by the most recent demographic data before the pandemic was declared. Table~\ref{tab:CountryUsersWealthLabels} reports the percentages of mobile phone users in our dataset divided into the three wealth groups that were considered in this analysis. More precisely, users living in high-wealth neighborhoods represent the wealthiest $20\%$ of the urban population and account, on average, for more than $45\%$ of the active user base. In contrast, users living in low-wealth neighborhoods represent the less-wealthy $40\%$ of the population and only account for $18\%$ of the mobile phone user base on average. More details about mobility indicators and the average number of users for each group changing behavior from the baseline period to the pandemic can be found in SI Sec.~10B. To provide a better measure of the size of the differences in the pre-pandemic and during-pandemic periods we provide raw numbers of individuals self-isolating, commuting and migrating (see SI Tab.1). Similarly, percentages of the share of people self-isolating at home and the share of people commuting can be found in SI Tab.2.
				
				\begin{table}[!ht]
					\centering
					\setlength{\tabcolsep}{5.pt}
					\begin{tabularx}{0.65\columnwidth}{@{}X rrrrrr@{}}
						\toprule
						\textbf{} & \textbf{BR} & \textbf{CO} & \textbf{MX} & \textbf{ID} & \textbf{PH} & \textbf{ZA} \\
						\midrule
						\textbf{\small{High wealth}} & 42.17 \% & 36.75 \% & 39.85 \% & 36.58 \% & 39.48 \% & 76.77 \% \\
						\textbf{\small{Med wealth}} & 34.75 \% & 41.51 \% & 38.87 \% & 44.95 \% & 39.91 \% & 18.3 \% \\
						\textbf{\small{Low wealth}} & 23.08 \% & 21.74 \% & 21.27 \% & 18.46 \% & 20.61 \% & 4.93 \% \\        \bottomrule
					\end{tabularx}
					
					\caption{Device distribution per wealth of home location neighborhood. For each country, the percentage of devices labeled within each different group is reported. A biased distribution is found, with users living in high-wealth neighborhoods being more represented than users living in low-wealth neighborhoods.}
					\label{tab:CountryUsersWealthLabels}
				\end{table}
				
				\subsubsection*{Commute distances}\label{sec:hw_distance_comparison}
				To compute distances between home and work locations in Fig.~\ref{fig:conditioned_model_parameters}, we leverage individual-level coordinates of home and workplaces and compute the Haversine distance between the two locations~\cite{Haversineformula}. We take the logarithm of distances to regularize the distribution. Then, log distances are standardized by subtracting the average of all users in a wealth group and dividing by the standard deviation, such that all distributions are centered around $0$ with a width of $1$.
				
				\subsection{Panel regression model}\label{sec:modeling_methods}
				We model the association between mobility indicators and policy restrictions for different socioeconomic groups. Mobility indicators by socioeconomic group are determined by the socioeconomic status of the home and work location of a user. 
				In this framework, the standardized mobility $mb_{ic}(t)$ of group $i$ in country $c$ at time $t$ is modelled as: 
				
				\begin{align*}
					mb_{ic}(t) = & a_i*incidence_g(t) + b_i*incidence_c(t)\\
					& + \sum_n c_{i,n}*C_{c,n}(t);
				\end{align*}
				where $a_i$ is a cross-country free parameter capturing the group-specific association with the global incidence of COVID-19 reported cases ($incidence_g(t)$), $b_i$ is the cross-country free parameter, pooled over all countries in our panel, to capture the association of the dependant variable with the local incidence of cases ($incidence_c(t)$), and $c_{i,n}$ are coefficients modeling the $C_n(t)$ association of policy $n$ with mobility for the $i$-group. Five different policy types are included in the model discussed in Sec.~\ref{sec:modeling_policy}: school closure policies, workplace closure policies, public transport closures, internal movement restrictions, and stay-at-home orders. $C_n(t)$ is the index associated with the $n$ policy at time $t$, as defined in ~\cite{hale2020oxford}, and $c_{i,n}$ is the corresponding group association coefficient. The $C_n(t)$ indices are a direct quantification of national and sub-nation stringency levels for the $n$ different studied policies (ranging from $0$, no policy in place, to $1$, maximum policy strength at full national level). A summary of the index computation procedure is reported in SI Sec. 12.A1~\cite{OxCGRT2022githubIndexDocumentation}. The choice of the components to include in the model is based on the hierarchical selection of the most important components in a modeling framework where policies are grouped in summary indices (containment, economic, and health)~\cite{hale2020oxford}. Model selection is performed using the Bayesian Information Criterion (BIC) comparing models with different covariate compositions. Selection robustness tests, performed on single countries as well as on single wealth-groups BIC values, provide consistent results in terms of model selection. Multicollinearity tests are performed in terms of the Variance Inflation Factor (VIF) and only models that do not present any factor with $VIF\ge4$ are included among those to be selected (see Sec. SI~10A)~\cite{james2013introduction}.
				The estimated coefficients reported in the main manuscript are those referring to the policy covariates and are estimated on the period starting from April 11th, 2020, and spanning until January 1st, 2021. Extensive robustness testing is performed to ensure the results' reliability. In Sec. SI~10B, we show parameter estimates for the different policy covariates performing the regression i) over a different time window, and ii) excluding from the regression one country at a time (to ensure stability over country selection). We refer to Sec. SI~10 for more details on the model. 
				It is important to stress that the adopted framework accounts for inter-country and inter-wealth-group differences using both standardized covariates and standardized target variables for each country and wealth group. This approach aims at finding differences in wealth groups' responses relative to each group's mobility patterns and thus focuses on modeling the relative response to incidence and/or policy enactment rather than the metrics differences as presented in Fig. 2, 3, and 4.

				\section*{Author contributions}
				L.L. and S.F. conceived the original idea and planned the experiments.
				A.C., A.M. provided geographical and demographic information.
				L.L., O.L.C., L.C., and L.M. processed the mobility data and carried out the experiments.
				L.L., O.L.C., B.L., N.L.G, and S.F. contributed to the interpretation of the results and wrote the manuscript.
				
				\section*{Competing Interests}
				The authors declare no competing interests.
				
				\section*{Acknowledgements}
				L.L. thanks G.K. for the insightful discussions and his support during the entire project development.
				L.L. has been supported by the ERC project ``IMMUNE'' (Grant agreement ID: 101003183).  L.L. acknowledges the support from the ``Fondazione Romeo ed Enrica Invernizzi'' for the research activities of the 'Covid Crisis Lab' at Bocconi University.
				B. L. acknowledges the
				support of the PNRR ICSC National Research Centre for
				High Performance Computing, Big Data and Quantum
				Computing (CN00000013), under the NRRP MUR program funded by the NextGenerationEU.
				The findings, interpretations, and conclusions expressed in this paper are entirely those of the authors. They do not necessarily represent the views of the International Bank for Reconstruction and Development/World Bank and its affiliated organizations, or those of the Executive Directors of the World Bank or the governments they represent.
				
				 \section*{Supplementary Information}
				Supplementary Information in support to the main manuscript can be found at the following link: \url{https://doi.org/10.6084/m9.figshare.23639079.v1}. Code and data to reproduce the results are shared under CC BY-NC-ND 4.0 License at \url{https://doi.org/10.6084/m9.figshare.23639079.v1}.
				
				 \newpage
				 \section*{References}
				 \bibliography{copy_bib}

			\vspace{2cm}
			\noindent
			\footnotesize{
			Supplementary Information in support to the main manuscript can be found at the following link: \url{https://doi.org/10.6084/m9.figshare.23639079.v1}. Code and data to reproduce the results are shared under CC BY-NC-ND 4.0 License at \url{https://doi.org/10.6084/m9.figshare.23639079.v1}.
				
			\end{document}